\documentclass[runningheads]{llncs}

\usepackage{amssymb}    
\usepackage{bm}         
\usepackage{amsmath}
\usepackage[T1]{fontenc}
\usepackage{multirow}
\usepackage{authblk}
\usepackage{hyperref}

\usepackage{graphicx,verbatim}
\usepackage{subcaption}
\usepackage{pgffor} 
\usepackage{subcaption}
\titlerunning{$trAIce3D$: A Prompt-Driven Transformer Based U-Net}

\begin{document}
\title{$trAIce3D$: A Prompt-Driven Transformer Based U-Net for Semantic Segmentation of Microglial Cells from Large-Scale 3D Microscopy Images}
%

\author{
MohammadAmin Alamalhoda\inst{1} \and
Arsalan Firoozi\inst{1,2} \and
Alessandro Venturino\inst{1} \and
Sandra Siegert\inst{1}
}

\authorrunning{Alamalhoda et al.}

\institute{
Institute of Science and Technology Austria, Klosterneuburg, Austria\\
\email{Ssiegert@ista.ac.at} 
\and
Department of Electrical Engineering, Columbia University
}

    
\maketitle              
\begin{abstract}
The shape of a cell contains essential information about its function within the biological system. Segmenting these structures from large-scale 3D microscopy images is challenging, limiting clinical insights especially for microglia, immune-associated cells involved in neurodegenerative diseases. Existing segmentation methods mainly focus on cell bodies, struggle with overlapping structures, perform poorly on noisy images, require hyperparameter tuning for each new dataset, or rely on tedious semi-automated approaches. We introduce $trAIce3D$, a deep-learning architecture designed for precise microglia segmentation, capturing both somas and branches. It employs a two-stage approach: first, a 3D U-Net with vision transformers in the encoder detects somas using a sliding-window technique to cover the entire image. Then, the same architecture, enhanced with cross-attention blocks in skip connections, refines each soma and its branches by using soma coordinates as a prompt and a 3D window around the target cell as input. Training occurs in two phases: self-supervised \textit{Soma Segmentation}, followed by prompt-based \textit{Branch Segmentation}, leveraging pre-trained weights from the first phase. Trained and evaluated on a dataset of 41,230 microglial cells, $trAIce3D$ significantly improves segmentation accuracy and generalization, enabling scalable analysis of complex cellular morphologies. While optimized for microglia, its architecture can extend to other intricate cell types, such as neurons and astrocytes, broadening its impact on neurobiological research.


\end{abstract}

\section{Introduction}
Cells are commonly classified by shared characteristics including morphology, physiological properties, and transcriptional states\cite{ref1,ref2}. For example, in the brain, neural branching patterns critically influence their function \cite{ref2_1}. However, accurate 3D microscopy cell segmentation remains challenging, requiring extensive manual effort while being susceptible to operator bias, limiting clinical applications where subtle structural changes could serve as early disease markers \cite{ref_Neuron_tracing,ref_U_Net}.

Microglia, the brain’s immune cells, undergo morphological changes signaling altered homeostasis \cite{ref_3DUNet,ref_VNet}. Traditional segmentation methods, such as semi-automated tracing and classical image processing, struggle with complex branching patterns and microscopy image brightness/contrast variations within a cell, requiring advanced hyperparameter fine-tuning \cite{ref_UNet_nature}. While deep learning has improved soma segmentation, existing models frequently fail to accurately capture both soma and intricate branches \cite{ref8_microglia_characterization}, particularly in volumetric microscopy where overlapping structures complicate individual branch segmentation. Moreover, segmentation methods based on classical image processing techniques handle data variability poorly \cite{ref_microglial_3d_seg,ref_3dMorph}

We present $trAIce3D$, a novel two-stage deep learning framework utilizing prompt-based segmentation for accurate instance-level 3D microglia segmentation. Prompt-based methods guide object isolation in complex scenes and have recently powered advanced segmentation models \cite{ref_PointRend,ref_SAM}. $trAIce3D$ combines Convolutional Neural Networks (CNNs) and vision transformers, leveraging U-Net's capabilities in biomedical image segmentation \cite{ref_3DUNet} while addressing CNNs' limitations in capturing long-range dependencies through a hierarchical 3D vision transformer backbone \cite{ref_An_Image_is_Worth_16_Words,ref_Swin_Transformer,ref_TransBTS}.

The framework operates in two stages: First, a U-Net with 3D vision transformer encoder and CNN decoder detects soma positions via sliding-window approach. In the second stage, the detected soma locations serve as prompts to guide individual cell branch segmentation using the same architecture equipped with cross-attention blocks in its skip connections.

$trAIce3D$, trained and evaluated on 41,230 microglial cells with diverse morphologies, demonstrates robust performance in microglia soma and arbor segmentation. While focused on microglia, $trAIce3D$ can be applied to other cell types with similar morphological characteristics, expanding its potential impact on neurobiological and clinical (diagnostic) research.

\section{Methods}
\begin{figure}
\includegraphics[width=\textwidth]{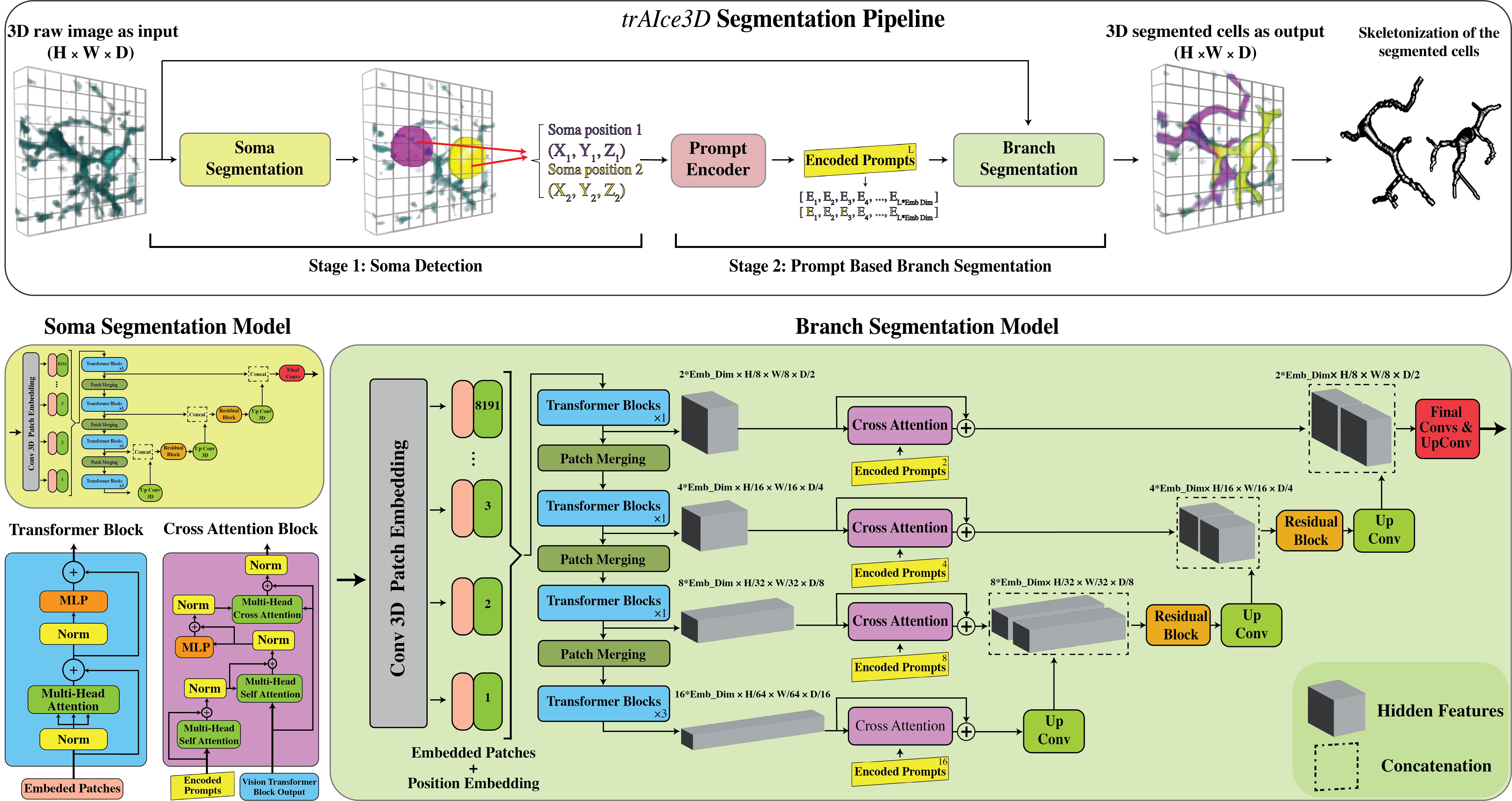}
\caption{Overview of the $trAIce3D$ segmentation pipeline and models architecture.} \label{fig1}
\label{fig:figure_1}
\end{figure}

$trAIce3D$'s pipeline (Figure \ref{fig:figure_1}) comprises two stages: Soma Detection and Branch Segmentation with Prompt-based Refinement. Both stages share the same encoder-decoder architecture (input/output size 256×256×16 pixels), with Branch Segmentation incorporating additional cross-attention blocks in skip connections for prompt integration. The framework was trained and evaluated on the 3D volumetric microscopy images containing labeled microglial cells from seven mouse brain regions covering development, adulthood, and neurodegeneration.  \cite{ref_morphomics}.

The full training pipeline and model architecture are publicly available at \href{https://github.com/AminAlam/trAIce3D}{Github.com/AminAlam/trAIce3D}.

\subsection{Dataset and Data Augmentation}
We utilized data from a published microglia morphology study \cite{ref_morphomics} containing 41,230 annotated cells from 230 mouse brain images. Microglial processes were traced using Imaris filament-tracing plugin v9.2 with manual verification.
Original images (median dimensions [2304, 2560, 56]) were standardized to [0.4, 0.4, 1.1] $\mu$m per pixel and divided into 256×256×16 pixel cubes, excluding those with <5\% microglia volume or discontinuous labels. Each cube underwent histogram equalization and (0,1) range normalization. Overlapping cubes were used for Soma Segmentation training, while cubes extracted around individual microglia were used for Branch Segmentation training.
TorchIO v0.20.4 \cite{ref_TorhcIO} provided spatial augmentations (RandomAffine, RandomFlip, RandomElasticDeform) and intensity augmentations (RandomNoise, RandomBlur, RandomGamma) during training to enhance generalization and model robustness.

\subsection{Architecture}  
The $trAIce3D$ framework consists of dual architectures for Soma and Branch Segmentation, sharing the same encoder-decoder infrastructure but utilize different skip connections. 
\begin{table}
\caption{Parameter distribution of $trAIce3D$ variants.}
\label{tab:parameter_distribution}
\centering

\begin{tabular}{|c|c|c|c|c|}
\hline
\multirow{2}{*}{Model} & \multirow{2}{*}{Total Parameters (M)} & \multicolumn{3}{c|}{Parameter Distribution (M)} \\
\cline{3-5}
 &  & Encoder & Skip Connections & Decoder \\
\hline
$trAIce3D_S$ & 6.10 & 2.98 & 0.42 & 2.70 \\
$trAIce3D_M$ & 27.38 & 11.88 & 4.69 & 10.81 \\
$trAIce3D_L$ & 109.39 & 47.45 & 18.72 & 43.22 \\
\hline
\end{tabular}
\end{table}
We developed three model variants with increasing computational capacity: $trAIce3D_S$ (Small), $trAIce3D_M$ (Medium), and $trAIce3D_L$ (Large), characterized by encoder vision transformer embedding dimensions of 32, 64, and 128, respectively. As both attention blocks in skip-connections and CNNs in the decoder are parameterized by the encoder's embedding dimension, larger dimensions systematically increase parameters across all model components, with distributions detailed in Table~\ref{tab:parameter_distribution}.

\subsubsection{Image Encoder (hierarchical 3D Vision Transformer).}
Our encoder employs a hierarchical 3D Vision Transformer to process volumetric microscopy images (256, 256, 16). The input is partitioned into (8, 8, 2) non-overlapping patches and embedded into either 32-, 64-, or 128-dimensional space ($trAIce3D$ variants S, M, and L respectively) using 3D convolution, with sinusoidal positional embeddings preserving spatial information. The architecture comprises six sequential transformer blocks, each featuring 8-head Flash Attention-based Multi-Head Self-Attention (MHSA) \cite{ref_FlashAtt} (Equation \ref{eq:MHSA}, where Q, K, and V are Queries, Keys, and Values respectively) and an Multi Layer Perceptron (MLP) with 8× hidden dimension and GeLU activation, incorporating layer normalization and residual connections.

\begin{equation}
\text{Attention}(Q,K,V) = \text{softmax}(\frac{QK^T}{\sqrt{d_k}})V
\label{eq:MHSA}
\end{equation}

Following the Swin Transformer Patch Merging approach \cite{ref_Swin_Transformer}, patch merging operations are performed after Transformer Blocks 1, 2, and 3. Each merging operation reduces spatial resolution by a factor of 8 while increasing the embedding dimension by a factor of 2 using a linear layer. This hierarchical structure progressively reduces the total spatial resolution by a factor of 512 ($8^3$) and expands the final embedding dimension by a factor of 8 from its initial value. Updated positional embeddings are maintained to ensure spatial consistency after each merging operation. The encoder outputs four hierarchical feature maps with shape $(H' \times W' \times D', \text{Emb\_Dim})$, where $H'$, $W'$, and $D'$ are the reduced spatial dimensions. These feature maps serve as skip connections to the decoder, enabling the capture of both local details and global context in volumetric microscopy analysis.

\subsubsection{Skip Connections.}
For Soma Segmentation, skip connections directly pass hierarchical encoder features to the decoder without modification. The Branch Segmentation model replaces traditional skip pathways with Residual Cross-Attention Modules (RCAM), enabling prompt-driven feature refinement.
\begin{equation}
\begin{aligned}
P' &= P + \text{Attention}(P + P_{pe}, P + P_{pe}, P) \\
P'' &= P' + \text{Attention}(I + I_{pe}, P' + P_{pe}, P') \\
P''' &= P'' + \text{MLP}(P'') \\
I' &= I + \text{Attention}(P''' + P_{pe}, I + I_{pe}, I) \\
\end{aligned}
\label{eq:cam}
\end{equation}
RCAMs integrate a two-way cross-attention mechanism with self-attention and MLP blocks. The mechanism has been shown in Equation \ref{eq:cam}, where $I$ represents image features from the encoder, $P$ encoded prompts, and $I'$ output of the RCAM. Each attention operation incorporates sinusoidal positional embeddings ($I_{pe}, P_{pe}$) and is followed by layer normalization. The module enables bidirectional interaction between image and prompt features while preserving spatial information through residual connections. This approach balances high-resolution detail with semantic awareness, particularly beneficial for complex branching structures and overlaying regions.

\subsubsection{Decoder.}
The decoder reconstructs high-resolution feature maps while preserving spatial and semantic details using residual convolutional blocks and skip connections. It uses a hierarchical feature-refinement process, integrating both low-level spatial and high-level semantic features for precise segmentation. Transposed 3D convolutions double the resolution and reduce channels by four at each stage. Up-sampled maps are concatenated with skip connections, retaining hierarchical information. A residual block with 3D convolutions (kernel 3×3×3, padding 1×1×1) maintains spatial consistency. Each layer includes batch normalization, GeLU activation, and dropout. The final segmentation map is refined by series of 3D convolutions and transposed, and finally passed through a sigmoid function.

\subsubsection{Prompt Encoder.}
The Prompt Encoder transforms spatial prompts into feature embeddings compatible with cross-attention modules. Given a set of point-based spatial cues \( P = \{p_1, p_2, \dots, p_N\} \) where each \( p_i = (x_i, y_i, z_i) \) represents a target soma location, the encoding process begins with normalization: \( \tilde{p}_i = \left( \frac{x_i}{W}, \frac{y_i}{H}, \frac{z_i}{D} \right) \). The normalized coordinates are projected into a high-dimensional space using a learned Gaussian matrix \( \mathbf{\Phi} \in \mathbb{R}^{3 \times d} \), yielding \( \mathbf{e}_i = 2\pi \cdot \tilde{p}_i \mathbf{\Phi} \). Fourier feature mapping is applied using sine and cosine functions \cite{ref_att_all_need}: \( \mathbf{PE}(p_i) = \left[ \sin(\mathbf{e}_i), \cos(\mathbf{e}_i) \right] \in \mathbb{R}^{2d} \), preserving spatial relationships. Each point \( p_i \) is assigned a learnable embedding \( \mathbf{E}_i = \mathbf{W}_{\text{point}} + \mathbf{PE}(p_i) \), where \( \mathbf{W}_{\text{point}} \in \mathbb{R}^{2d} \) is a trainable parameter. The final feature representation for a batch of \( N \) points is \( \mathbf{E} = [\mathbf{E}_1, \mathbf{E}_2, \dots, \mathbf{E}_N] \in \mathbb{R}^{N \times 2d} \).

\subsection{Training and Computational Efficiency}
\subsubsection{Training.}
Training proceeds in two stages: first training Soma Segmentation on 3D microscopy images with binary soma masks, then transferring these weights to initialize Branch Segmentation's encoder and decoder, which accepts additional soma coordinate inputs. Models trained for 200 epochs (batch size 64, gradient accumulation 2) using 1e-4 learning rate with cosine decay (50 warmup epochs, final 1e-6) on two NVIDIA A100 GPUs using PyTorch 2.6.0 \cite{ref_pytorch} + cu12.6 with BF16 precision.
Soma Segmentation uses a balanced loss (0.5×Focal Loss+0.5×Dice Loss), while Branch Segmentation employs an optimized combination (0.2×Focal Loss+0.6×Dice Loss+0.2×clDice Loss). Focal Loss \cite{ref_focalLoss} addresses class imbalance:
\begin{equation}
FL(p_t) = -\alpha (1 - p_t)^\gamma \log(p_t)
\label{eq:focal}
\end{equation}
with $\alpha=0.25$, $\gamma=3$, and $p_t$ is the model's estimated probability for the true class. clDice Loss \cite{ref_clDiceLoss} enhances tubular structure segmentation:
\begin{equation}
\mathcal{L}_{clDice} = 1 - \frac{2 \sum S T}{\sum S + \sum T}
\label{eq:cldice}
\end{equation}
where $S$ and $T$ are predicted and ground truth skeletons, respectively.

\subsubsection{Computational Efficiency.}
For each model variant, we measured the average inference GPU memory usage and speed per volumetric tile (256 × 256 × 16) on an NVIDIA A100 (80GB VRAM). The $trAIce3D$ variants S, M, and L require 0.5, 2, and 6 GB and run at about 90, 50, and 30 volumes/min, respectively. These metrics help users balance computational efficiency and accuracy when selecting a model.

\section{Experiments and Results}
\subsection{Experiments}
\subsubsection{Soma Segmentation (80/15/5 Split).}
We benchmarked $trAIce3D$ Soma Segmentation against CellPose cyto3 \cite{ref_cellPose} and nnU-Net 3D Cascade Full Resolution \cite{ref_nnUnet} as two widely used frameworks for cell segmentation. nnU-Net uses a seven-stage PlainConvUNet with InstanceNorm3D, LeakyReLU, Z-score normalization, processing (256, 256, 16) patches (batch size 64, 17.8M parameters). CellPose predicts cell probability maps and vector flow fields from microscopy datasets. With an 80/15/5 train/test/validation split, we trained all models except CellPose (inference-only due to limited 3D training support) for 200 epochs with data augmentation. Evaluation focused on binary soma detection—classifying locations as "segmented" or "unsegmented"—quantified via accuracy, F1-score, and precision.


\begin{figure}[!htbp]
\includegraphics[width=\textwidth]{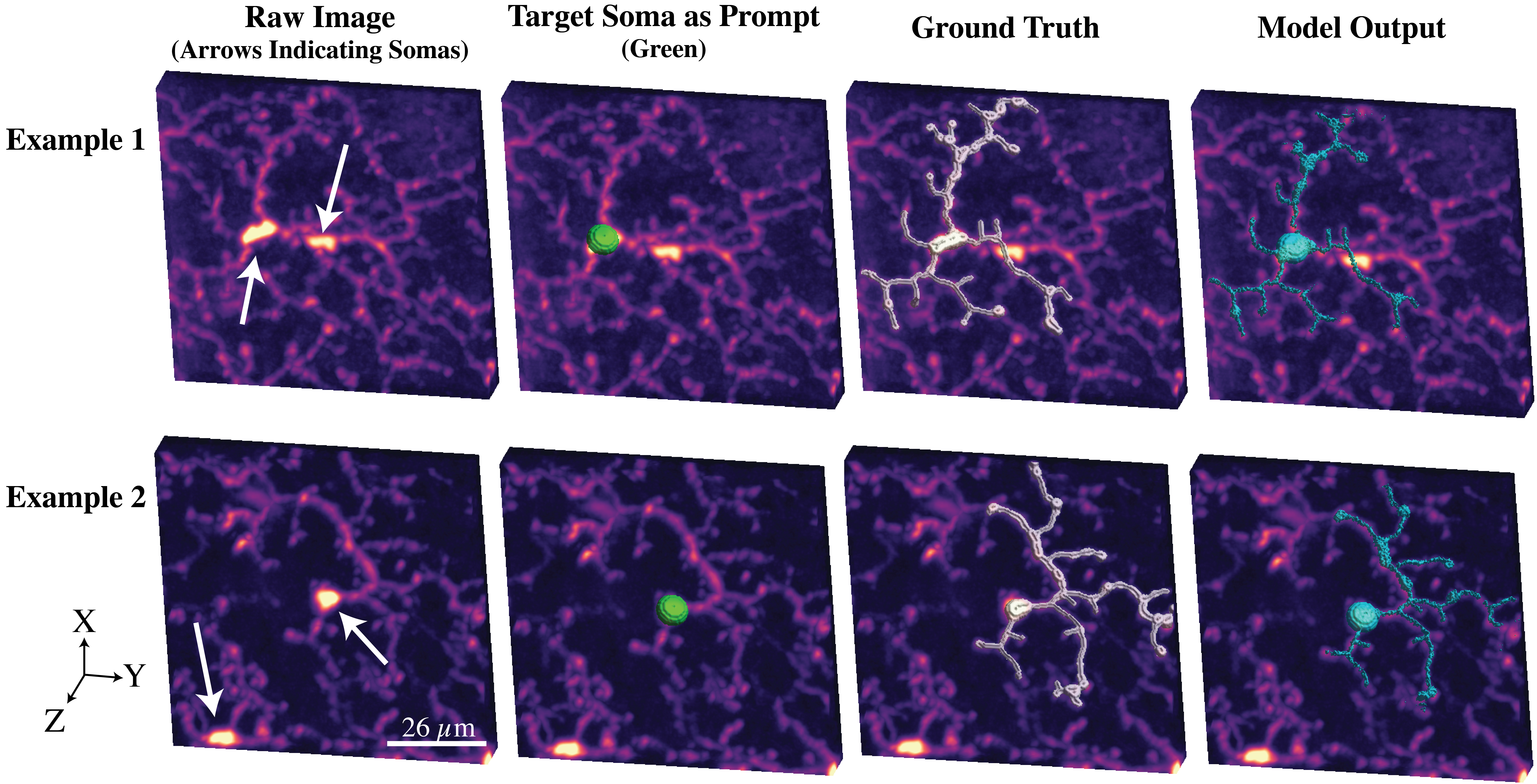}
\caption{Examples of qualitative results of $trAIce3D_L$.} \label{fig1}
\label{fig:figure_2}
\end{figure}

\subsubsection{Branch Segmentation (Model Reuse + 5-Fold Cross-Validation).}
For branch-segmentation, we evaluated $trAIce3D$ variants (S, M, and L) exclusively, as no publicly available models, automatic pipelines, or large datasets exist for 3D microglia instance-level segmentation. We initialized encoder and decoder weights of each variant from our pretrained soma segmentation models and trained them for 200 additional epochs with data augmentation, monitoring for overfitting.

To validate $trAIce3D_L$‘s ability to capture complex branching structures, we conducted 5-fold cross-validation for both segmentation tasks. Each fold’s Branch Segmentation model was initialized with encoder/decoder weights from its corresponding Soma Segmentation training. Performance was evaluated using three complementary metrics: Dice Score (DS), Average Path Length Difference (APLD), and Hausdorff Distance (HD)—measuring spatial accuracy and branching topology.

\subsection{Results}
The performance of $trAIce3D$ was evaluated both qualitatively and quantitatively. Figure \ref{fig:figure_2} illustrates its accurate segmentation of branching structures, even in complex cases with overlapping cells in 3D space. 

Table \ref{tab:detection_branch_results} shows that $trAIce3D_M$ and $trAIce3D_L$ outperform CellPose and nnU-Net in Soma Detection, with $trAIce3D_L$ achieving an F1-score of 87.5\% compared to nnU-Net’s 64\%. Even the smallest model, $trAIce3D_S$, performs better than CellPose (F1 Score 56.8\% vs 39\%) and slightly worse than nn-Unet (F1 Score 56.8\% vs 64\%) despite having significantly fewer parameters. 

\begin{table}
\caption{Soma and Branch Segmentation models performance results.}
\label{tab:detection_branch_results}
\centering
\begin{tabular}{|l|c|c|c|c|c|c|}
\hline
\multirow{2}{*}{Model Name} & \multicolumn{3}{c|}{Soma Segmentation} & \multicolumn{3}{c|}{Branch Segmentation} \\
\cline{2-7}
 & Accuracy & F1 Score & Precision & DS & APLD & HD \\
\hline
CellPose & 34.9\% & 39\% & 36\% & - & - & - \\
nn-Unet & 60.5\% & 64.0\% & 61.7\% & - & - & - \\
$trAIce3D_S$ & 51\% & 56.8\% & 53.5\% & 0.31 & 2.90  & 16.53\\
$trAIce3D_M$ & 78.5\% & 81.6\% & 80.3\% & 0.47 & 1.04 & 9.22 \\
$trAIce3D_L$ & \textbf{83.1\%} & \textbf{87.5\%} & \textbf{84.3\%} & \textbf{0.63} & \textbf{0.77} & \textbf{5.1} \\
\hline
\end{tabular}
\end{table}

For Branch Segmentation, $trAIce3D_L$ achieved the highest performance across all metrics, including a Dice Similarity (DS) of 0.63, an Average Path Length Difference (APLD) of 0.77, and a Hausdorff Distance (HD) of 5.1 (Table \ref{tab:cross_validation}). The medium-sized model ($trAIce3D_M$) also performed good, while $trAIce3D_S$ struggled with segmentation of finer structures which resulted in poor performance.

The five-fold cross-validation (Table \ref{tab:cross_validation}) confirms the robustness of the $trAIce3D_L$ model. For Soma Detection, it achieves an 85.2\% F1 score, indicating high accuracy. In Branch Segmentation, it attains a Dice score of 0.61, APLD of 0.91, and HD of 5.99, demonstrating its ability to capture intricate branching structures. These results emphasize the importance of model size and embedding dimensionality in representing complex cellular morphologies and highlight the effectiveness of vision transformers in segmenting branching structures, especially when trained on large datasets.

\begin{table}
\caption{5-fold cross validation results of $trAIce3D_L$.}
\label{tab:cross_validation}
\centering
\begin{tabular}{|c|c|c|c|c|c|c|c|}
\hline
\multirow{2}{*}{Fold} & \multicolumn{3}{c|}{Soma Segmentation} & \multicolumn{3}{c|}{Branch Segmentation} \\
\cline{2-7}
 & Accuracy & F1 Score & Precision & DS & APLD & HD \\
\hline
1 & 81\% & 85\% & 82\% & 0.53 & 1.49 & 7.23 \\
2 & 79\% & 83\% & 80\% & 0.64 & 0.75 & 5.5 \\
3 & 82\% & 87\% & 85\% & 0.68 & 0.69 & 5.31 \\
4 & 77\% & 82\% & 79\% & 0.56 & 0.92 & 6.49 \\
5 & 84\% & 89\% & 87\% & 0.65 & 0.72 & 5.42 \\
\hline
Average (±std) & \textbf{80.6}±2.7\% & \textbf{85.2}±2.9\% & \textbf{82.6}±3.4\%\ & \textbf{0.61}±0.06 & \textbf{0.91}±0.29 & \textbf{5.99}±0.56 \\
\hline
\end{tabular}
\end{table}

\section{Conclusion}

We introduced $trAIce3D$, a prompt-driven transformer-based U-Net for instance-level segmentation of microglial cell somas and branches in large 3D microscopy images. By integrating vision transformers in the encoder and cross-attention in skip connections, $trAIce3D$ effectively incorporates prompt-based guidance for precise microglia segmentation.

Our results show that $trAIce3D$’s Soma Segmentation model outperforms CNN-based models like nnU-Net, and 5-fold cross-validation demonstrates the robust performance of its Branch Segmentation model. We also highlight the critical role of embedding dimensionality in capturing intricate microglial morphologies. A key novelty is the use of cross-attention in skip connections, enabling accurate single-cell segmentation by integrating target cell soma location as prompt information into feature refinement.

Despite these achievements, limitations remain. Performance drops under varying imaging conditions, and new microscopy modalities may require fine-tuning. Scaling to larger structures (e.g., neurons) requires adaptive window-shifting, while high cell densities hinder segmentation due to low SNR. Future work could explore Shifted Window Attention, hybrid vision transformer-CNN models \cite{ref_Swin_Transformer,ref_SwinUnet3D} for better feature representation, and deep learning-based window-shifting for larger structures.

Overall, $trAIce3D$ offers an efficient, scalable, and high-accuracy solution for 3D cell segmentation in large-scale microscopy datasets, which could advance biological and clinical (diagnostic) applications.

%
%
%
%

\end{document}